\documentclass[12pt,preprint]{aastex}

\begin{document}
\title{Detection of polarization from the $E^4\Pi-A^4\Pi$ system of F\lowercase{e}H in sunspot spectra}

\author{A. Asensio Ramos, J. Trujillo Bueno\altaffilmark{1}, M. Collados}
\affil{Instituto de Astrof\'{\i}sica de Canarias, 38200, La Laguna, Spain}
\altaffiltext{1}{Consejo Superior de Investigaciones Cient\'{\i}ficas, Spain}
\email{aasensio@ll.iac.es, jtb@ll.iac.es, mcv@ll.iac.es}

\begin{abstract}
Here we report the first detection of polarization signals
induced by the Zeeman effect in spectral lines of the $E^4\Pi-A^4\Pi$ system of FeH
located around 1.6 $\mu$m. Motivated by the tentative detection of this band in the intensity spectrum of late-type dwarfs, we have
investigated the full Stokes sunspot spectrum finding
circular and linear polarization signatures that we associate
with the FeH lines of the $E^4\Pi-A^4\Pi$ band system.
We investigate the Zeeman effect in these molecular transitions
pointing out that in Hund's case (a) coupling the effective Land\'e factors
are never negative. For this reason, the fact that our spectropolarimetric observations
indicate that the Land\'e factors of pairs of FeH lines have opposite signs,
prompt us to conclude that the $E^4\Pi-A^4\Pi$ system must be in intermediate
angular momentum coupling between Hund's cases (a) and (b).
We emphasize that theoretical and/or laboratory investigations
of this molecular system are urgently needed for exploiting its promising
diagnostic capabilities.

\end{abstract}


\section{Introduction}
FeH constitutes one of the most important opacity contributors in late-type dwarfs, in the red and near-infrared between 0.7 $\mu$m and
1.3 $\mu$m. It was however detected in the atmospheres of late M dwarfs much later than many of the other hydrides formed with less
abundant atomic species. This is probably due to the fact that the FeH spectrum is very complicated, arising
from quartet and sextet terms (Langhoff \& Bauschlicher 1990). Although some bands are in the optical and blue part of the spectrum,
the crowding of atomic lines makes it difficult to distinguish FeH bands. For this reason, the most studied
FeH electronic system is that produced by the transition $F^4\Delta-X^4\Delta$. This band system, widely used in studies of
late-type dwarfs, produces a conspicuous absorption near 1 $\mu$m.
A theoretical analysis of this band system has been performed by
Phillips et al (1987) with the assignation of quantum numbers to many of the observed lines of the $v=0-0$ band.

In a recent study of the infrared intensity spectrum of sunspots, Wallace \& Hinkle (2001) identified almost 70 lines
between 1.58 $\mu$m and 1.755 $\mu$m common to both the sunspot spectrum and a furnace laboratory spectrum of FeH . They tentatively
associated these lines to the
$E^4\Pi-A^4\Pi$ system based on the theoretical work of Langhoff \& Bauschlicher (1990). These authors predicted this band system to be
around 2
times weaker than the $F^4\Delta-X^4\Delta$ system, even though the $E^4\Pi-A^4 \Pi$ is one of the strongest bands of the quartet system
in FeH. Later, Cushing et al (2003)
compared the near-infrared spectrum of four late-type dwarfs with the laboratory FeH spectrum, finding 34 features that dominate in the
$H$-band spectra. They associated some of these features to the $v=0-0$ $E^4\Pi-A^4\Pi$ band of FeH. They found a very
similar behavior of this band and the other IR bands of FeH when observing stars of different spectral types, thus reinforcing that these
features belong to FeH.

There are almost no studies of the polarization properties of FeH lines. A first attempt has been carried out by Berdyugina et al (2001)
and by Berdyugina \& Solanki (2002) for lines of the $F^4\Delta-X^4\Delta$ system, assuming that the angular momentum coupling is that
given by Hund's case (a) (see Herzberg 1950). These authors were forced to use Hund's
case (a) coupling because no estimation of the spin-orbit coupling constants of the electronic states of the transition are available.
In view of the effective Land\'e factors, they concluded that the $F^4\Delta-X^4\Delta$ band system
of FeH might be of interest for the investigation of the magnetic properties of solar and stellar atmospheres.

In this letter, we present the first full Stokes observations of FeH
in the near infrared, showing
that the $E^4\Pi-A^4\Pi$ band system must be in
intermediate coupling between Hund's cases (a) and (b).
To our knowledge, this is the first time polarization signals in FeH are
observationally detected in sunspots.

\section{Observations}
The observations were carried out on June 7, 2002 with the Tenerife Infrared Polarimeter (TIP; see Mart\'{\i}nez Pillet et al 1999)
mounted on the German Vacuum Tower Telescope (VTT) at the Observatorio del Teide (Spain). The observed sunspot was located out of
the solar disk center, at $\mu=0.68$ (being $\mu$ the cosine of the
heliocentric angle), so that linear polarization signals
may be expected. The total size of the umbra was $\sim$18". The presence of a
light bridge crossing the sunspot umbra led us to select only those points
within the umbra whose polarization properties are not contaminated
by the presence of the light bridge.
Interestingly, the depth of the observed FeH intensity profiles
is reduced close to the light bridge, possibly due to the dissociation
of the FeH molecules caused by a temperature increase.
Similarly, we also found a smaller amplitude in the Stokes $V$ spectrum.

The spectral resolution of the observation was $\sim$26 m\AA\, with a total wavelength coverage of $\sim$7 \AA. In order to investigate
the polarization properties of FeH lines, we performed three scannings with
a step of 0.4" for different spectral regions and three time series with the slit crossing the center of the umbra. The integration
time for each position in the scannings was 1 second, while the total integration time for the time series was between 5 and 10 minutes.
Although the detection of the FeH features is obtained also in the
1 second integrations, we have used the temporally averaged Stokes
profiles because they have better signal-to-noise ratio.
The typical noise level is $\sim10^{-4}$ of the continuum intensity.

As shown below in Fig. 4, the observed FeH lines,
apart from producing conspicuous antisymmetric $V$ signals,
also show perfectly detectable symmetric Stokes $Q$ and $U$ profiles.
However, in this first paper we will
only focus on Stokes $V$.

\section{The Zeeman effect in F\lowercase{e}H}
Since no perturbation analysis has been performed for any of the electronic states of FeH, no spin-orbit coupling constants are available.
This makes it necessary to treat the lines of both the $F^4\Delta-X^4\Delta$ and the $E^4\Pi-A^4\Pi$ system in any of the limiting
Hund's coupling cases.
However, according to Berdyugina \& Solanki (2002), the $F^4\Delta-X^4\Delta$ system is known to be in intermediate
coupling between (a) and (b) and strong deviations
of the effective Land\'e factor $\bar g$ for the lines of the P and R branches are expected from that given by Hund's case (a). In
particular, $\bar g$ for the lines of the P and R branches will increase as $J$ increases.
This increase in the Zeeman sensitivity
has been observed by Wallace et al (1999) in the intensity spectrum of sunspots associated to an increase in the broadening of the
high $J$ lines. A similar behavior is expected for the lines of the
$E^4\Pi-A^4\Pi$ band. Wallace \& Hinkle (2001) apparently detected such a behavior from
the line splitting in the intensity spectrum. They found many FeH lines which appear undoubled but a set of seven lines around 15930 \AA\
which present a splitting, probably caused by the Zeeman effect since it increases with the field strength in sunspots.

For those electronic levels with $\Lambda \neq 0$, the spin-orbit coupling constants are usually large (see Huber \& Herzberg 2003).
Therefore,
the energy separation associated to the
multiplet splitting is expected to be large. We also expect the spin-orbit
coupling to be large enough so that the field at which the
transition to the Paschen-Back regime occurs is
larger than the typical field strength of sunspots. In fact, this is the
case with all the
molecules studied by Berdyugina \& Solanki (2002) presenting $\Lambda \neq 0$. If the FeH lines that belong to the $E^4\Pi-A^4\Pi$
system are indeed in the Zeeman regime, we expect Stokes profiles similar to those of a normal Zeeman triplet. This is in fact
confirmed by the
observations.

\begin{figure}
\plotone{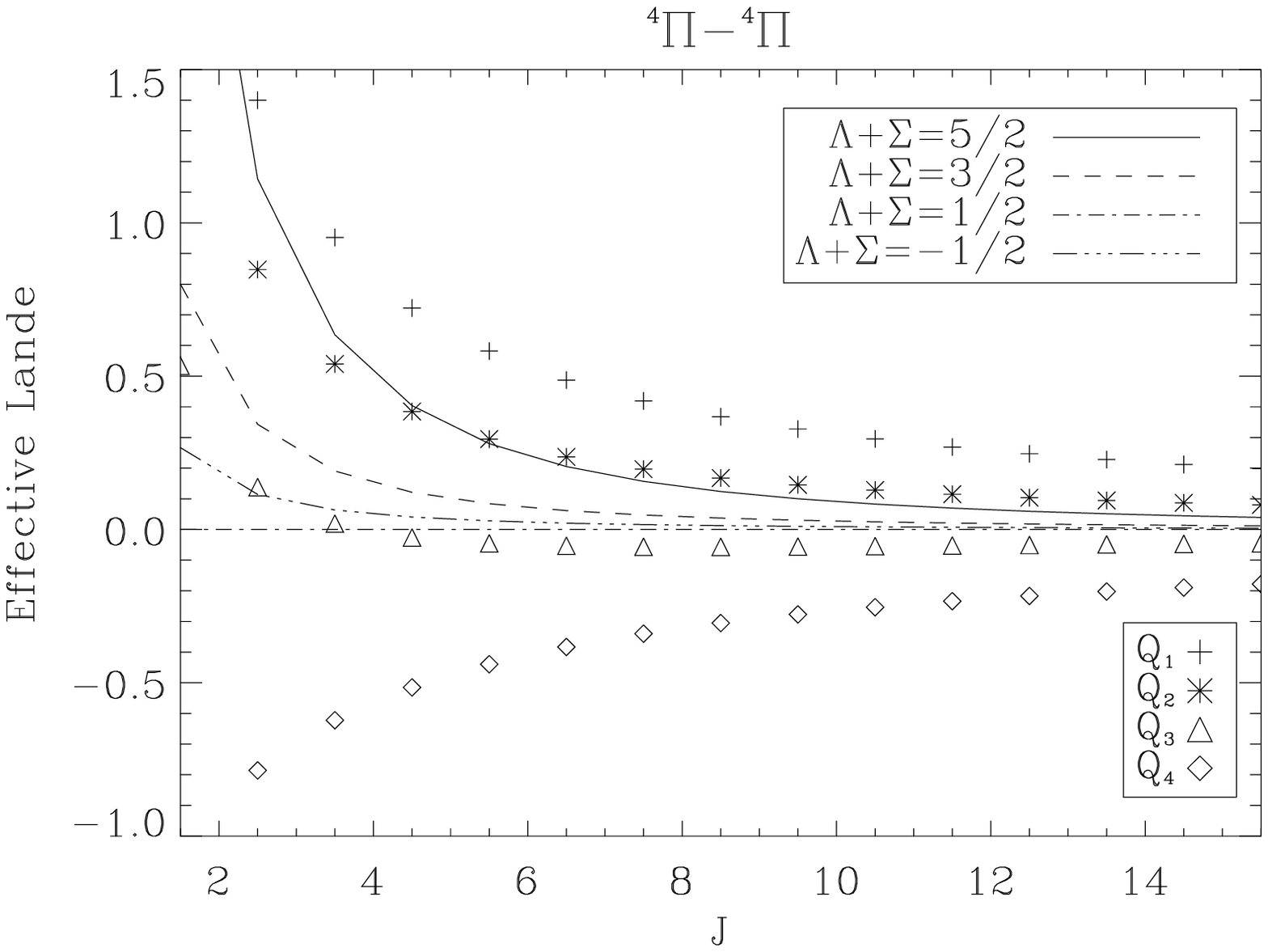}
\caption{Effective Land\'e factor of the Q branch of the $E^4\Pi-A^4\Pi$ band
system of FeH calculated using Hund's case (a)
coupling (lines) and Hund's case (b) coupling (symbols).
The lines between the $\Lambda+\Sigma=1/2$ levels are completely insensitive to the magnetic
field in Hund's coupling case (a). Note that while the effective Lande factor is always non-negative for Hund's case (a) coupling, it can
become negative for Hund's case (b) coupling.\label{fig_lande_FeH}}
\end{figure}

The effective Land\'e factors
calculated in Hund's case (a) coupling for the Q branch lines
of the
$E^4\Pi-A^4\Pi$ band are shown in Fig. \ref{fig_lande_FeH}.
The effective Land\'e factor ($\bar g$) can be obtained easily from
Herzberg's  (1950) formulae for the Land\'e factors of the lower and upper levels.
Since such Q-branch lines result
from transitions between electronic levels with the same quantum numbers,
$\bar g$ can be written simply as
\begin{equation}
\bar g = \frac{(\Lambda+2\Sigma)(\Lambda+\Sigma)}{J(J+1)},
\end{equation}
where $\Lambda=1$ and $\Sigma=3/2,1/2,-1/2,-3/2$. We point out that
$\bar g=0$ for the lines of the P and R branches and that the lines between levels with $\Lambda+\Sigma=1/2$ are insensitive to the magnetic field. The rest of the spectral
lines belonging to transitions between low $J$ levels seem to be as sensitive to the magnetic field as those of the $F^4\Delta-X^4\Delta$
system studied by Berdyugina \& Solanki (2002). It is very
important to note that all the values of $\bar g$ are non-negative, contrary to
what happens for the $F^4\Delta-X^4\Delta$ system. We also show in Fig. \ref{fig_lande_FeH} the effective Land\'e factor obtained for
the lines of the main Q branch using Hund's case (b) coupling. Note that in this coupling case $\bar g$ can take positive and negative values. The same is valid for the lines of the P and R branches.

Although little magnetic field diagnostics can be presently done with these lines given the lack of precise spectroscopic data,
we have used the observed polarization signals in order to obtain a first insight into the potential diagnostic interest of these FeH lines.

\section{Discussion}
We have observed three spectral regions. A region around 16605 \AA, another one around 16110 \AA, and a third
one around 16575 \AA. The observation of the first region was motivated by
the presence of two vibration-rotation OH lines of the $X^2 \Pi$ level, similar to those observed by Harvey (1985). Because the
spectroscopic constants of OH are known,
these lines can be used to obtain information about the magnetic field in sunspots.
To this end, as shown in Fig. 2, we have performed LTE syntheses in the hot
umbra model of Collados et al (1994)
finding that  a constant vertical magnetic field of 1800 G pointing radially outwards
leads to a fairly good fit to the observed Stokes profiles in
the OH lines\footnote{The LTE synthesis in the cool umbra model of
Collados et al (1994) presents OH line absorptions much deeper than the observations.}.
The Zeeman patterns for the OH lines have been obtained applying the theory
developed by Schadee (1978), thus they do not depend on
any assumption about the coupling.
We point out that since the OH lines are stronger than the FeH lines, a magnetic
strength of 1800 G gives only a lower limit to the strength of the field in the deeper regions
of the sunspot umbra where the FeH polarization is originated.
Assuming a typical gradient of 5 G km$^{-1}$ (Collados et al 1994),
we estimate that the magnetic field in the formation region of the FeH lines may be
about 600 G higher than in the OH formation region.
For this reason, we have assumed ${\rm B}{\approx}2400$ G in the following estimations of the effective Land\'e factors of the observed FeH lines.

In order to obtain information about the approximate value
of the effective Land\'e factor of the FeH lines, we have applied two different techniques. The first one assumes that the
line is formed under the weak field (WF) approximation (e.g. Landi Degl'Innocenti 1992).
Molecular lines usually have small effective
Land\'e factors except for lines that arise between levels with very small values of $J$ (see Herzberg 1950;
Landau \& Lifshitz 1982). Since the Zeeman splitting in the IR is
relatively large, and assuming typical thermal and microturbulent velocities, the weak field approximation is valid for
$\bar g B \ll 1000$ G, which is somewhat restrictive for some of our FeH lines. The second technique assumes that the line is formed in the
strong field (SF) regime so that peak separation in Stokes $V$ is indicative of the separation of the $\sigma$ components when they are
fully split. Since the lines are not in any of these limiting regimes, the correct effective Land\'e factors will be between the values obtained via the two techniques (assuming that our estimation of the sunspot magnetic field strength is sufficiently afortunate).

Fig. \ref{fig_FeH_TIP_2} shows two FeH lines at 16108 \AA, one of which produces a clear circular polarization signal, while the
other is apparently insensitive to the Zeeman effect. We show in the same figure the fit obtained to the line at 16108.3 \AA\
by plotting the derivative of the intensity spectrum multiplied by the scaling factor using $\bar g_\mathrm{WF} =0.09$,
which represents an approximation to the effective Land\'e factor of this line. Note that the fit of the magnetic sensitive
line is strikingly good for this combination
of field strength and $\bar g$. From the previous fit,
we can infer that the strong field regime has not been reached, since the
separation of both Stokes $V$ lobes are correctly obtained with the weak field formula.
This is reinforced by the low value obtained for $\bar g$. In the strong field regime, we obtain
$\bar g_{\mathrm{SF}} \approx 0.24$.

The FeH line at 16107.8 \AA\ may be assumed to have $\bar g \approx 0$,
in principle, since at first sight there seems to be
no significant signal in Fig. 3. However, a closer inspection indicates that a very small
$V$ signal with $\bar g < 0$ seems to be present in our observations.
This may constitute a first indication
that deviations from Hund's case (a) are expected for these lines.
Another but much stronger indication will be discussed below.

Fig. \ref{fig_FeH_TIP_3} shows three FeH lines around 16575 \AA\ which have been detected
previously in intensity by Wallace \& Hinkle (2001). These lines
are quite weak in the sunspot intensity spectrum
and it has been a bit difficult to apply the weak-field approximation since the derivative of the
intensity profile turns out to be very noisy. We have obtained $\bar g_\mathrm{WF} \approx 0.38$
for the line
at 16571.5 \AA\ and $\bar g_\mathrm{WF} \approx -0.38$
for the line at 16576.8 \AA. According to the results shown in
Fig. \ref{fig_lande_FeH},
a negative $\bar g$ is not possible in Hund's case (a) coupling.
This constitutes our strongest proof that this band of FeH is in
the intermediate coupling case between Hund's cases (a) and (b).
The coupling constants and the line identifications have to be obtained
in case the lines of this band are to be used as tools for diagnosing
magnetic properties of the coolest regions of sunspot atmospheres
or magnetic fields on cool dwarfs.
The line at 16575 \AA\ seems to be a blend of some FeH lines at
slightly different wavelengths due to the intricate structure of Stokes $V$
with apparent cancellations among them. The
linear polarization signal is shown in the lower panel of Fig. \ref{fig_FeH_TIP_3}.
Note that all the FeH lines produce conspicuous
Stokes $Q$ and $U$ features with an amplitude of around 1\%.
If we assume that the lines are in the strong field regime, we obtain
$\bar g_\mathrm{SF} \approx 0.39$
for the line at
16571.5 \AA\ and $\bar g_\mathrm{SF} \approx -0.28$ for the
line at 16576.8 \AA\ from the peak separation
in Stokes $V$.

Our previous results can be compared with the effective Land\'e factor obtained from the
fully split FeH lines in the
atlas of the umbral spectrum of Wallace \& Livingston (1992).
For a field of 3500 G obtained from the fully split V {\sc i}  line
at 16570.5 \AA\ whose effective Land\'e factor is $\bar g=0.7$,
one obtains $\bar g_{\mathrm{SF}} \approx 0.27$ and $\bar g_{\mathrm{SF}} \approx -0.19$
for the lines at 16571.5 and 16576.8 \AA\,
respectively. Comparing with the results obtained directly from our spectropolarimetric
observations, we verify that in the sunspot we observed we are in the intermediate field regime.

\section{Conclusions}
We have shown observational evidence that the lines of the $E^4\Pi-A^4\Pi$ electronic system of FeH present circular and linear polarization
signals that are produced
by the Zeeman effect.
We have estimated the magnetic field strength with the help of two OH lines and inferred
the effective Land\'e factors of the FeH transitions using either the weak-field or the strong-field approximations. We have shown that
the $E^4\Pi-A^4\Pi$ band of FeH presents lines with negative effective Land\'e factors,
which is impossible under Hund's case (a) coupling.
Therefore, we conclude that this molecular band system must be in intermediate
angular momentum coupling between Hund's cases (a) and (b). The FeH lines studied here are
potentially interesting for empirical investigations of the physical conditions
in the lower atmosphere of sunspots and of the magnetism of late-type dwarfs.
To this end, theoretical and/or laboratory investigations
of this FeH molecular system are urgently needed.

\acknowledgments

We are grateful to Svetlana Berdyugina for suggesting useful improvements
to the original version of this letter. Thanks are also due to Egidio Landi Degl'Innocenti
for his careful reading of our paper.
This work has been partially supported by the Spanish Ministero de Ciencia y Tecnolog\'{\i}a through project AYA2001-1649.

%%%%%%%%%%%%%%%%%%%%%%%%%%%%%%%%%%%%%%%%%%%%%%%%%%%%%%%%%%%%%%%%%
% The bibliography
%%%%%%%%%%%%%%%%%%%%%%%%%%%%%%%%%%%%%%%%%%%%%%%%%%%%%%%%%%%%%%%%%

\begin{figure}
\plottwo{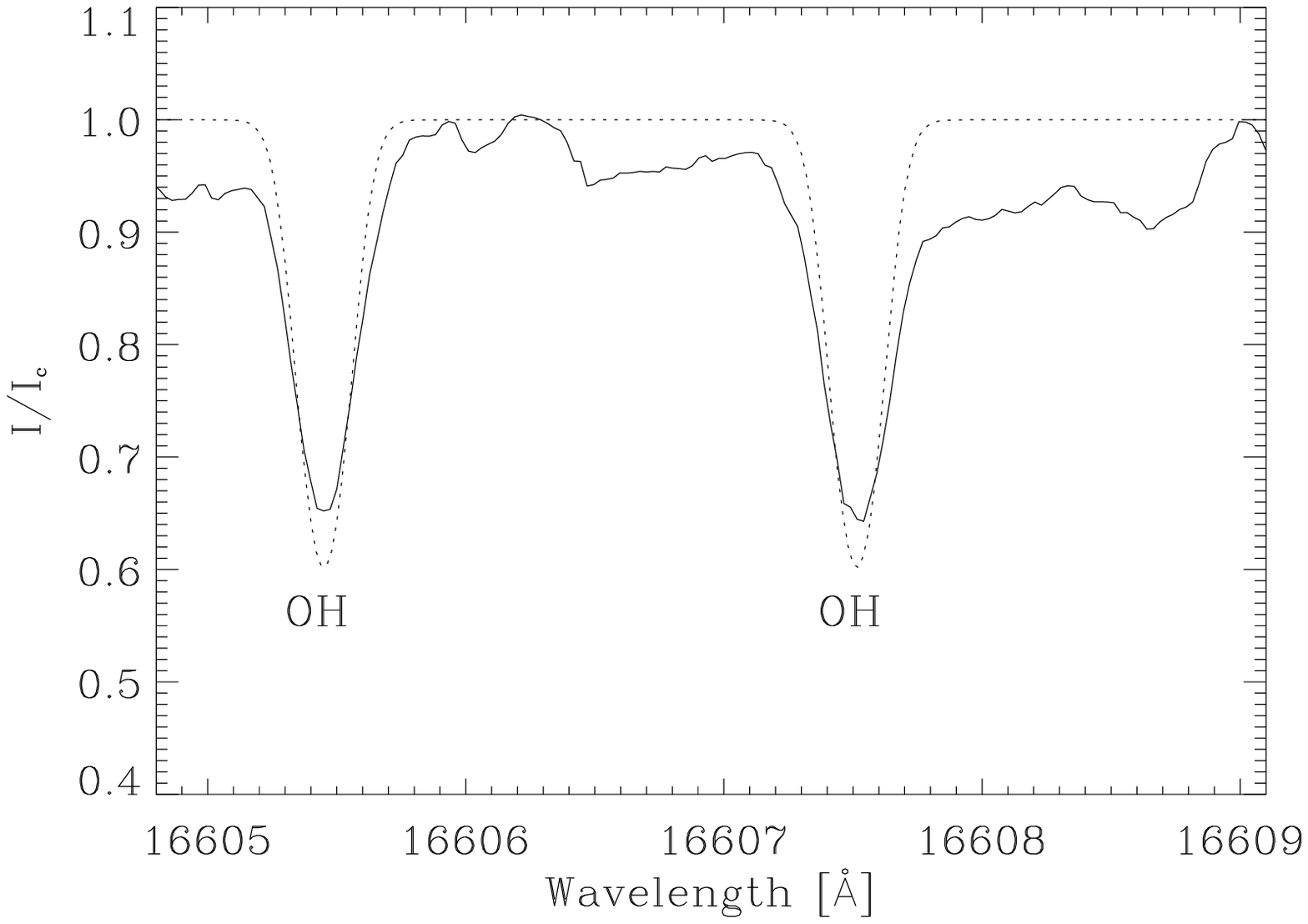}{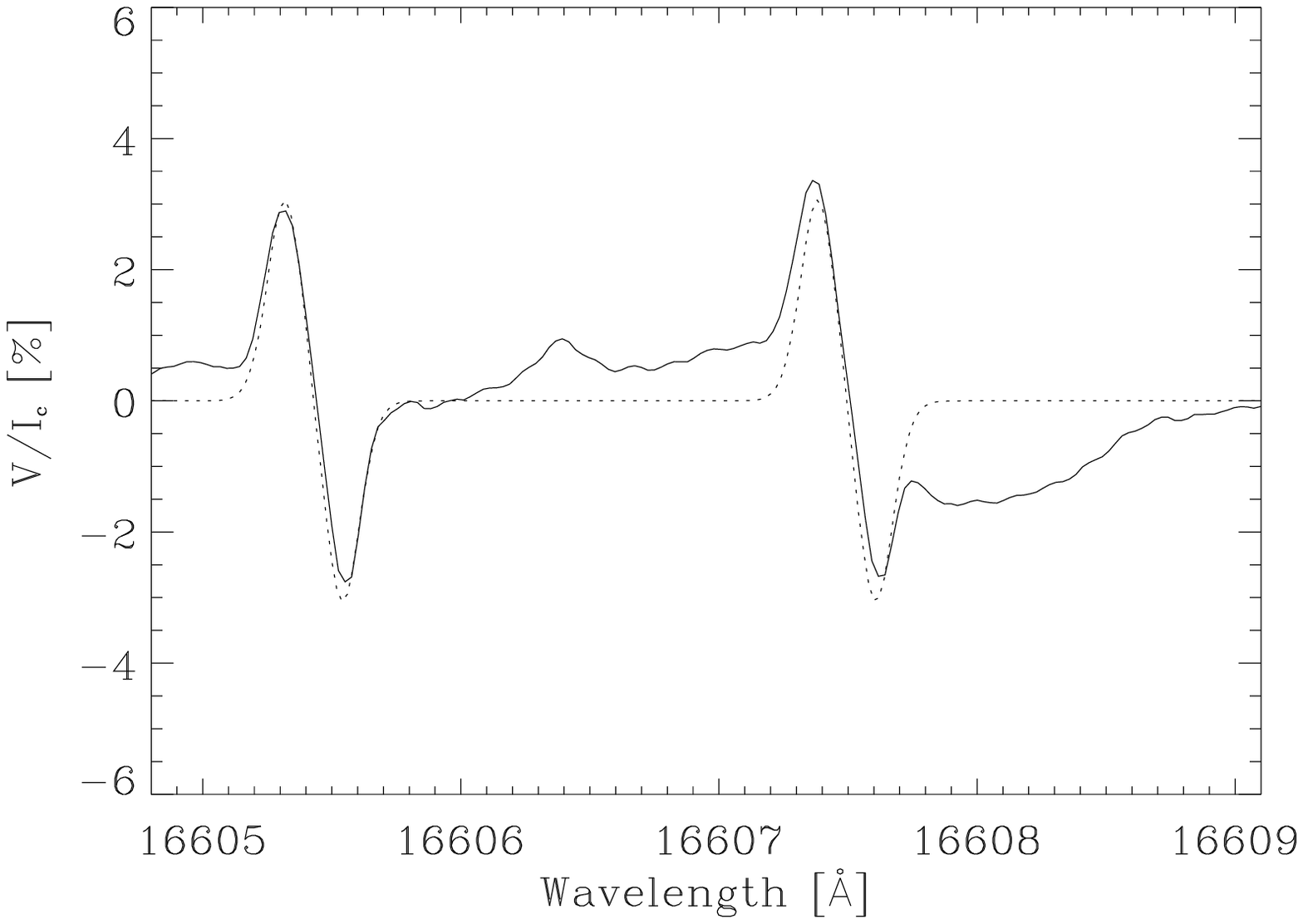}
\caption{Two OH lines close to the studied FeH lines. Since the coupling constants of OH are known,
we have calculated the emergent Stokes profiles
from the Collados et al (1994) hot umbra model, finding that a constant magnetic field of 1800 G
provides a fairly good
fit the observed Stokes profiles. The solid line represents the observation while
the dotted line is the modeling.\label{fig_FeH_TIP_1}}
\end{figure}

\begin{figure}
\plottwo{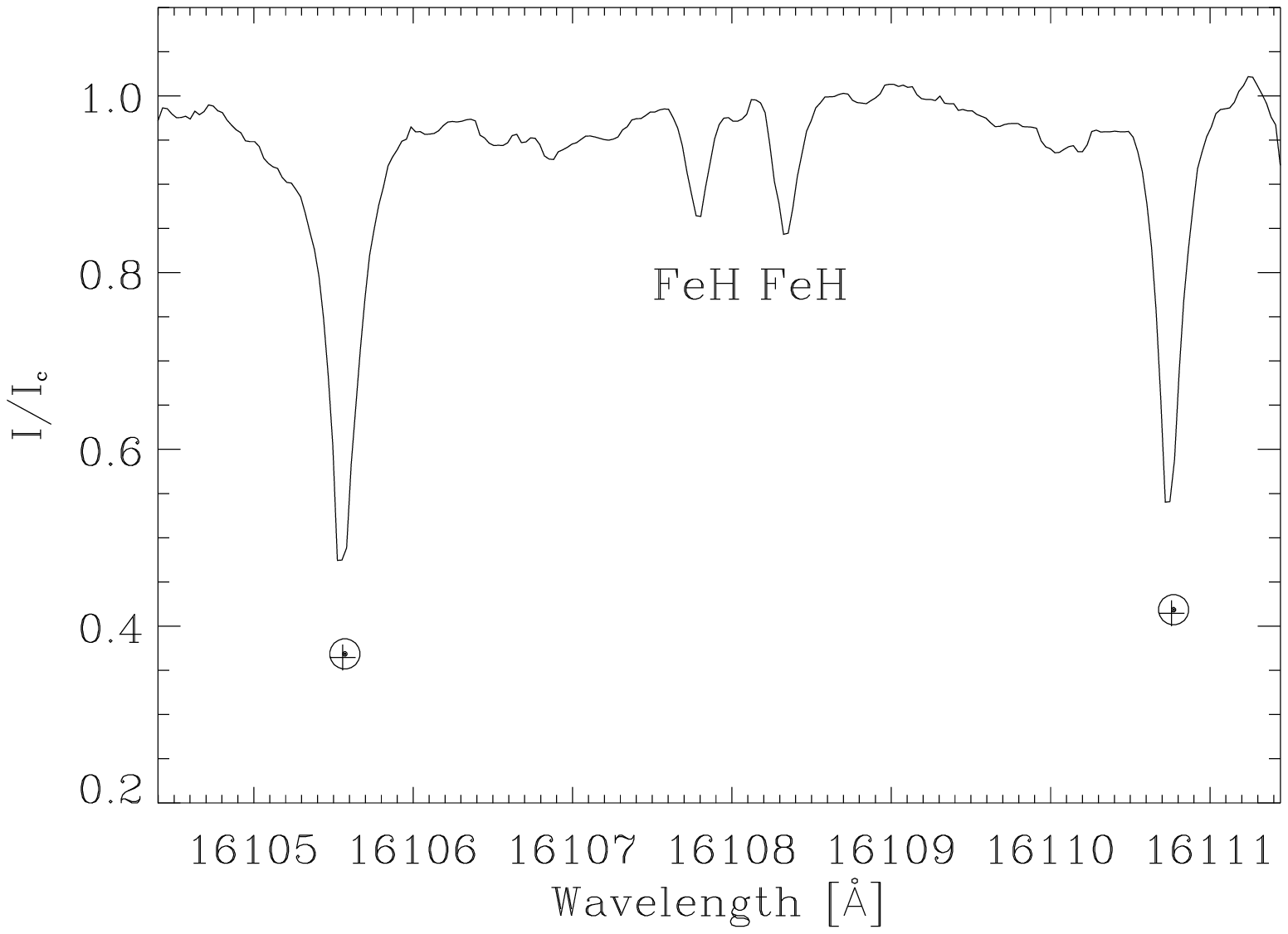}{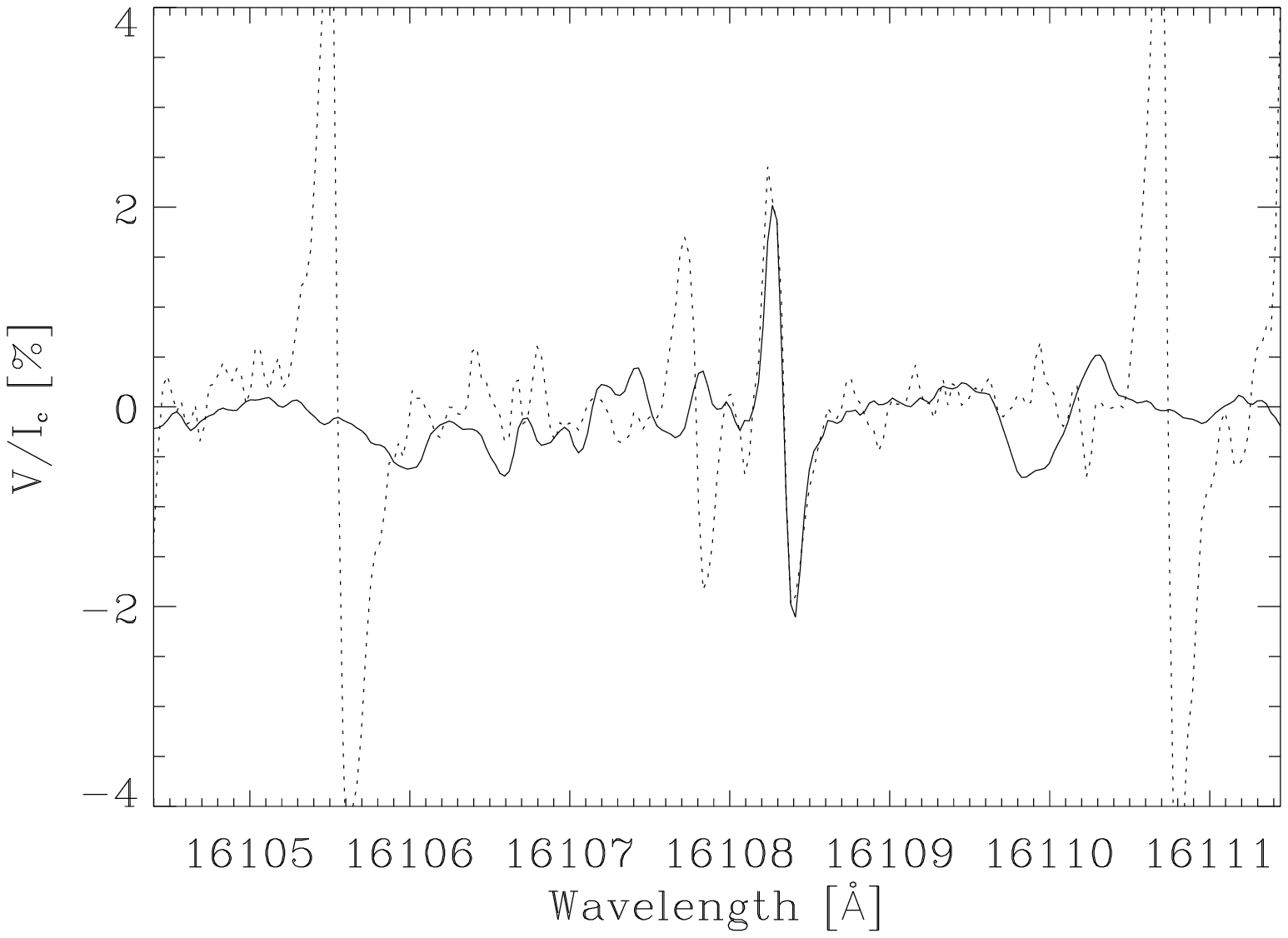}
\caption{Observed FeH lines together with the fit obtained via the weak-field
approximation for $B=$2400 G.
The two strong lines in
the extremes
of the spectral interval are telluric absorption produced by CH$_4$.
The solid line represents the observation while the dotted line
the derivative of the intensity spectrum
scaled with $\bar g=0.09$ in
order to fit the line at 16108.3 \AA.\label{fig_FeH_TIP_2}}
\end{figure}

\begin{figure}
\plottwo{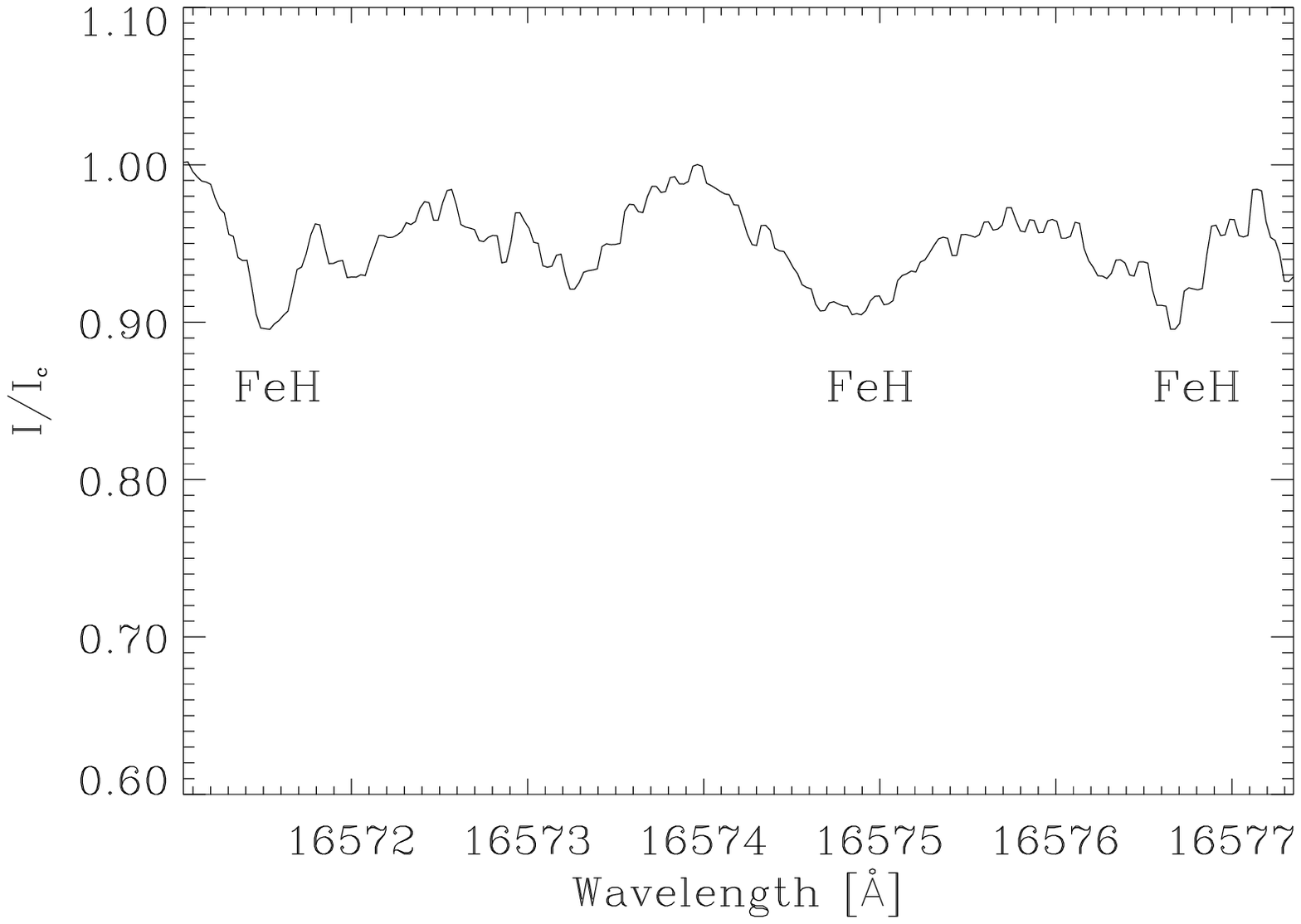}{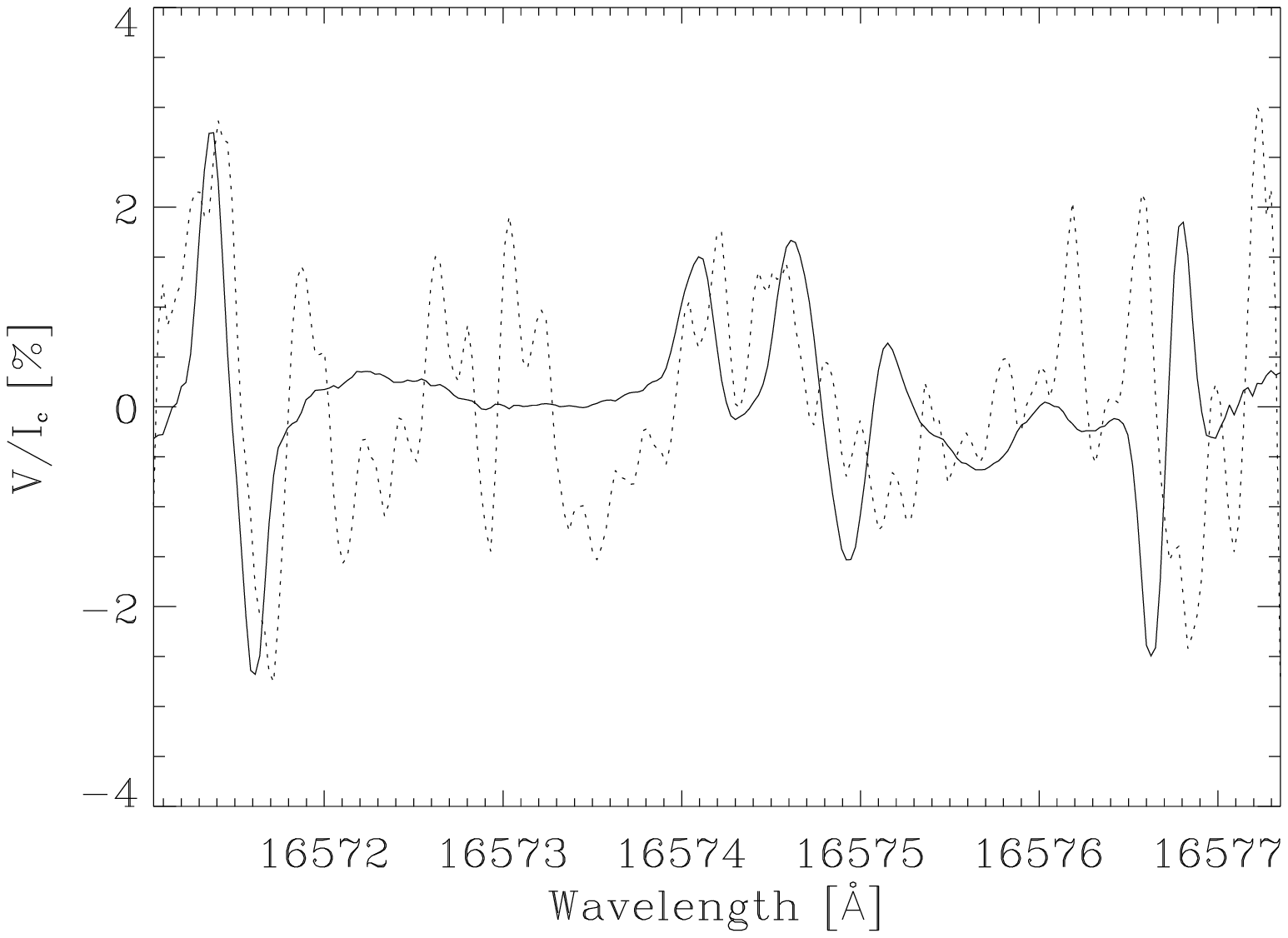}
\epsscale{2.2}
\plottwo{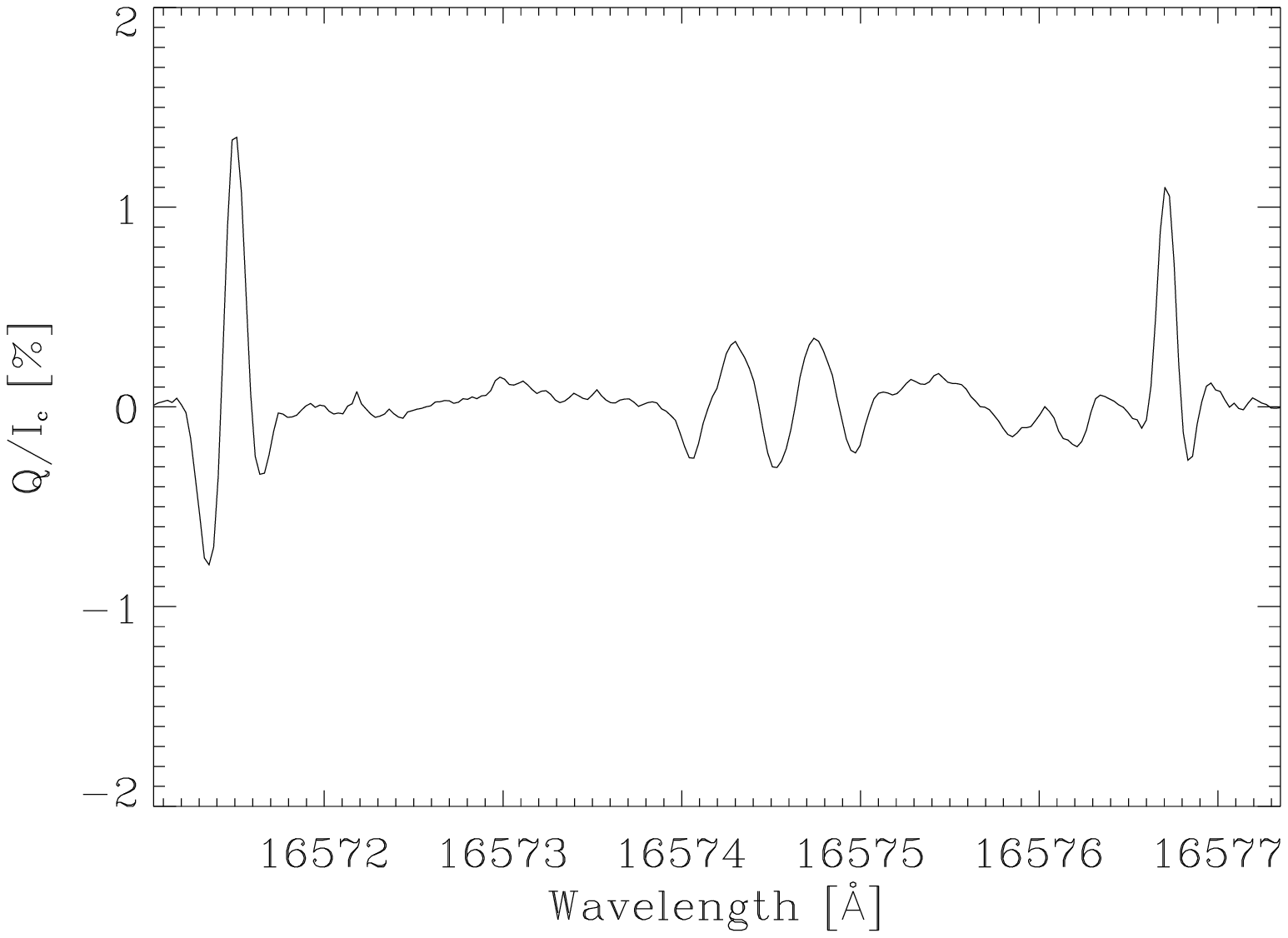}{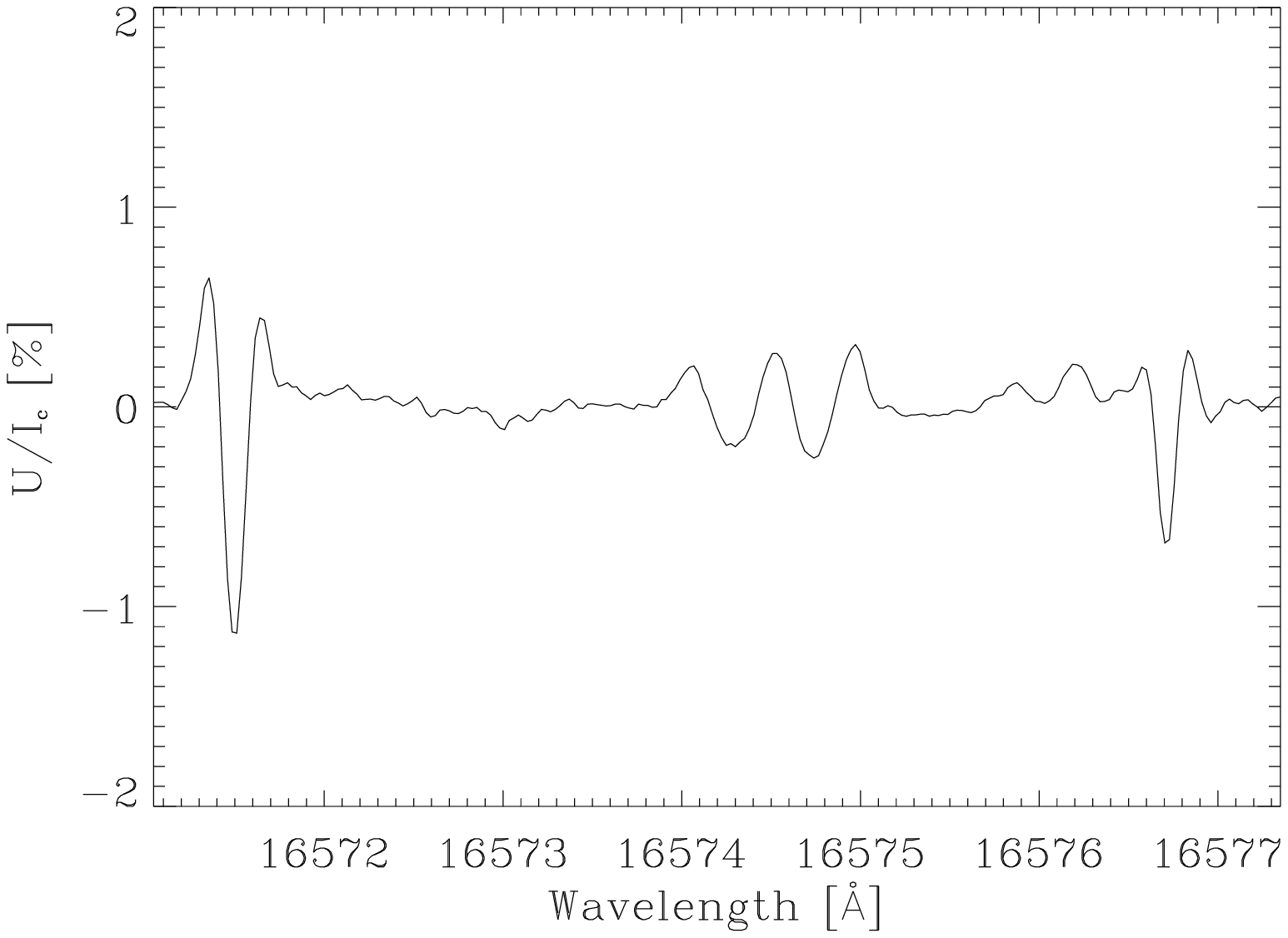}
\caption{Region with several FeH lines producing weak absorptions
in the intensity spectrum but very clear features in
the circular polarization spectrum. We show also the derivative of the intensity spectrum scaled
with $\bar g=0.38$. Note that the
line near 16577 \AA\ has negative effective Land\'e factor. The two lower panels show
the observed linear polarization spectrum.\label{fig_FeH_TIP_3}}
\end{figure}

\end{document}